# Structure of Quasi-One Dimensional Ribbon Colloid Suspensions

Thomas R. Stratton[a], Sergey Novikov[a], Ream Qato[a], Sebastian Villarreal[a], Bianxiao Cui[b], Stuart A. Rice[a,*] and Binhua Lin[a,c,*]

[a]Department of Chemistry, [a]The James Franck Institute and [c]CARS,

The University of Chicago, Chicago, IL 60637

[b]Department of Chemistry, Stanford University, Stanford, CA 94305

**Abstract**

We report the results of an experimental study of a colloid fluid confined to a quasi-one dimensional (q1D) ribbon channel as a function of channel width and colloid density. Our findings confirm the principal predictions of previous theoretical studies of such systems. These are (1) that the density distribution of the liquid transverse to the ribbon channel exhibits stratification; (2) that even at the highest density the order along the strata, as measured by the longitudinal pair correlation function, is characteristic of a liquid; and (3) the q1D pair correlation functions in different strata exhibits anisotropic behavior, resembling that found in a Monte Carlo simulation for the in-plane pair correlation function of a hard sphere fluid in a planar slit.

* corresponding authors: s-rice@uchicago.edu; blin@uchicago.edu

# I. Introduction

Interest in the equilibrium and transport properties of confined dense fluids has grown considerably in the last two decades. This interest has been driven in part by fundamental questions concerning the influence of dimensionality on the properties of a dense fluid and in part by the practical applications of confined fluids. The majority of the theoretical and experimental studies have confined attention to either quasi-one-dimensional (q1D) or quasi-two-dimensional (q2D) systems. This paper is concerned with systems that are, in a sense, intermediate between q1D and q2D. We report the results of an experimental study of the changes in the equilibrium structure of a colloid fluid that accompany the first stage of transition from q1D to q2D confinement, specifically the structure of q1D ribbons as a function of ribbon width.

There are a variety of structures associated with a confined fluid, and these structures are dependent on the geometry of the confinement. It has been known for some time that a three-dimensional (3D) hard sphere fluid in contact with a smooth hard wall has a density distribution along the normal to the wall that is stratified over a distance of several hard sphere diameters [1-4]. The length scale for the stratification depends on the liquid density and can include many layers in the high-density limit [5]. Recent experimental studies of confinement-induced layering in q1D systems focused on particles interacting via a screened Coulomb potential [6], quasi-2D dusty-plasma liquids [7] and repulsive magneto-rheological colloids in two-dimensional channels [8, 9]. These systems all have only purely repulsive pair interactions.

When a hard sphere liquid is confined to a q2D geometry by parallel walls only a few hard sphere diameters apart, thereby forming a slit, the density distributions that are generated at each liquid-wall interface overlap and interfere. As expected from the one-wall-liquid interface structure, the predicted density distribution of the liquid along the normal to the parallel walls is stratified [10] with, depending on the wall separation, the locations and amplitudes of the peaks in density somewhat different from those in the isolated one-wall-liquid interface. Moreover, a q2D system confined by parallel smooth planar walls supports many different ordered solids, the equilibrium structures of which depend on the wall spacing [11]. Specifically, the system supports the transitions *fluid* → 1Δ → 2□ → 2Δ → ... , where the symbol Δ represents triangular (hexagonal) packing, □ represents square packing and the numerical label is the number of layers. Alternation of square and triangular packing in the ordered solid persists for at least five layers. As to the liquid side of the transition line, a recent study has shown that, at any particular wall separation, there are ordered fluctuations in the liquid that anticipate the structure to which the q2D liquid freezes [12].

In a q1D ribbon system fluctuations destroy all possible order not imposed by the explicit symmetry breaking associated with the boundary conditions. The smooth walls that define a q1D ribbon channel break continuous rotational symmetry but support rotations by $\pm\pi$, and permit density modulation along the normal to the walls [9]. The length scale for the modulation depends on the liquid density and can include many layers in the high-density limit; the number of strata is dependent on the ribbon width. Simulations of hard spheres constrained to have q1D ribbon geometry confirm that even at very high density, and despite the existence of well-defined layering, the translational order in the system parallel to the walls resembles that of a liquid [13].

The results reported in this paper extend previous studies from this laboratory [14, 15]. Specifically, we report experimental studies of the structure of a colloid suspension confined to have a q1D ribbon geometry, as a function of ribbon width, focusing attention on changes in structure accompanying the transition from q1D to q2D confinement. The questions addressed include: (1) What is the nature of the density distribution transverse to the long axis of the q1D ribbon channel, and how does it change with colloid density and ribbon width? (2) What is the nature of the pair correlation function along the long axis of the ribbon channel in the different strata of the colloid liquid, and how does it depend on ribbon width?

## II. Experimental Details

Our basic experimental system consists of a sample cell that contains an aqueous suspension of colloidal silica spheres confined to a long narrow channel. Because it is very difficult to load the sample cell with a specified colloid density, precise replication of any particular sample density is not commonly achieved. Instead, many cells were used to generate samples with colloid densities close to the target density. Overall verification of the observations was generated by comparison of samples with similar but not identical colloid densities.

The channels we used were printed on a polydimethysiloxane substrate from a master pattern fabricated lithographically on a silicon wafer (Stanford Nanofabrication Facility, Stanford, Calif.). To prepare the sample cell a drop of colloid suspension was enclosed between the polymer mold and a cover slip with a polymer spacer (~100 μm), so that the top of the groove was open to a layer of fluid. The cover slip, supported on spacers, limits the rate of evaporation of the sample. The colloid particles settle gravitationally and fill the channel to an extent dependent on the bulk colloid suspension density. Previous studies in this laboratory have shown that the vertical motion of the colloid particles in the channel is very small [16]. Other studies [14] that use the same sample of colloid particles establish that the small residual charge on the colloid particles does not make a measurable contribution to the colloid-colloid effective interaction (it is independent of the ionic strength of the supporting fluid).

Figure 1 displays a schematic of the experimental setup. Digital video microscopy was used to extract time-dependent two-dimensional trajectories of the colloid particles (time resolution 0.033 s). We studied samples with fixed values of the colloid diameter, $\sigma$, and channel depth, $h$, for different channel widths, $w$, and colloid packing fraction. Further details of sample preparation and data analysis have been described elsewhere [14].

We have used two different samples of colloid particles. In one set of experiments the particle diameter was $\sigma = 1.58 \pm 0.04$ μm (density 2.2 g/cm$^3$, Duke Scientific, Fremont, Calif.), the channel depth was $3 \pm 0.2$ μm, and each sample was examined over a field of view of 108 μm diameter within a 2 mm long channel. Because of the extension of the channel beyond the field of view examined these samples can be treated as having open channel ends. The channel is also open to particle exchange with the bulk supernatant fluid. We found that the 1.58 μm spheres could not fill the channel with a packing fraction greater than 0.5. To generate samples with packing fractions greater then 0.5 we used 3.01 μm spheres (density 1.960 g/cm$^3$, Bangs Laboratories, Fishers, Ind.) in channels with widths that duplicate the ratios $w/\sigma$ of the experiments with the 1.58 μm colloids (see Table 1). Figure 2 displays some snapshots of instantaneous particle configurations for our ribbon channel systems. The number of particles inside the field of view varied from 23 to 450 per image dependent on the colloid packing fraction and channel width. We used the sphere center determination technique described by Crocker and Grier [17] to create histograms of the experimental data and to plot the trajectories of individual colloid particles, examples of which are provided in Figure 3. The influence of an optical artifact arising from overlap of particle images on the inferred effective colloid-colloid interaction is discussed in the Appendix to this paper.

We note that at equilibrium the non-uniform distribution of colloid particles in the channel has uniform chemical potential equal to that of the bulk colloid suspension, by virtue of the possibility for exchange of particles between the channel and the supernatant suspension, and that the gravitational contribution to the chemical potential is sufficiently large that the supernatant suspension is dilute. It is the difference in the gravitational contributions to the chemical potentials of the two colloids with different diameters (and masses) that permits generation of larger packing fraction with the larger particle.

From histograms of the particle locations we verified that the particle density distributions in all of our samples were uniform along the channel axis (the $x$-axis). This observation permits us to infer that there are no ripples or other structures in the channel walls that would induce order in the liquid parallel to the walls or contribute to an apparent colloid–colloid interaction. Examination of our images reveals that the extent of particle displacement transverse to the channel (the $y$-direction) is less than expected from the channel width and the absence of direct colloid-wall interaction other than excluded volume [14]. We attribute the enhanced confinement of a colloid particle in the $y$-direction to the non-wetting of the silicone elastomer wall by the aqueous colloid suspension, which creates a gap between the surface of the liquid and the channel wall. The effective width of the channels is reduced by approximately 1 μm along each wall largely irrespective of $w$ but dependent on $\sigma$. The effective width of the channel was calculated by generating the colloid density distribution transverse to the channel from a histogram of the colloid particle positions, and measurement of the distance between the most widely separated density peak centers; the effective channel width, $w_{eff}$, is this peak separation plus one particle diameter, $\sigma$, hence the precision for $w_{eff}$ is the same as that of the particle diameters. However, for clarity, we will refer to experimental data sets by the relevant fabricated channel width, i.e. "the 5 µm channel" as opposed to "the 3.1 µm channel". The packing fraction was calculated from the effective width using $\phi = N\pi(\sigma/2)^2 / lw_{eff}$, where $l$ is the length of the channel section in the field of view ($l$ = 108 µm), and $N$ is the average number of spheres in the section during the period studied. Because of the way the cells are filled it is not possible to precisely predetermine the concentration of colloid particles, so comparisons of data sets are for comparable but not identical packing fractions.

The pair correlation function of the colloid particles for a homogeneous q2D system can be calculated from the digital video microscopy data using

$$g_2(r) = \phi^{-1}\left\langle \sum_i \sum_j \delta(r_i)\delta(r_j - r)\right\rangle = \frac{N(r)}{2\phi\Delta r},$$

in which $N(r)$ is the number of colloid particles located within $\Delta r/2$ from the reference particle, $\Delta r$ is the step size of the histogram plot, and $\phi$ is the packing fraction of the colloid particles. Similarly, $g_2(x)$, the pair correlation function for a homogenous q1D system, can be calculated from the above expression, simply replacing $r$ and $\phi$ with $x$ and $\eta$ ($\eta$ is the q1D line packing fraction; $\eta = N\sigma/l$).

Where appropriate, we have calculated both $g_2(r)$ for all the particles in the channel and, $g_2(x)$ for each individual stratum in the transverse density distribution. In the case of $g_2(x)$, $N$ includes all the particles belonging to the same stratum whose width is determined by the two minima on either sides of the peak (see Figs 4-6). In the case of $g_2(r)$, it is worthwhile to note that, because of the geometry of the ribbon channel, the normalization $g_2(r)$ is, for any $r$, defined in a rectangular domain. By definition, $g_2(r)$ is normalized with respect to the distribution of a pair of randomly placed points in the same volume/area. Thus, the normalization factor is a constant defined by the system length in a one-dimensional system, while for a two-dimensional system the normalization factor is proportional to $r^{-1}$ because the radial domain area, $2\pi r dr$, increases with increasing $r$. Rather than determine an analytic form for the normalization factor in the rectangular domain, it was determined via a simulation. Because the actual confinement of the colloid particles is tighter than the nominal channel width, the effective width of the channel was determined from the experimental data, from the minimum and maximum $y$-positions of the particles. Then a random number generator was used to distribute two points uniformly in the rectangular area, and a histogram of pair separations generated. This histogram defines the normalization factor to within a constant.

### III. Results

**A. Transverse Density Distribution**

The density distribution of the liquid transverse to the ribbon channel exhibits stratification, and the modulation of the strata depends on the packing fraction of the liquid and the width of the channels.

Figure 5 displays the transverse density distributions of 1.58 μm colloid particles with packing fraction close to 0.41 (the range is 0.406 – 0.440) in ribbon channels with nominal widths of 5 μm, 8 μm, 11 μm, 14 μm and 20 μm (the uncertainty of the channel width is about ±0.1µm). The ratios of effective channel width to particle diameter, $w_{eff}/\sigma$, for these cases have the values 2.0, 3.6, 5.7, 7.4 and 11.3 and the data clearly show that the liquid in the channel is stratified with 2, 3, 5, 6 and 9 layers, respectively. We note that for this packing fraction bulk liquid conditions are not reached until the channel has more than 6 strata. Because the non-wetting condition generates a softer wall than does hard sphere-hard wall repulsion, the density peaks adjacent to the walls are not as asymmetric as predicted from simulation studies of that system [1, 5]. The small asymmetry of the density distribution across the channel is likely due to very slight variation in the flatness of the surface of that particular molded channel.

As expected, relative to the system with packing fraction $\phi \approx 0.41$ shown in Fig. 5, the modulation of the transverse density distribution is smaller and the widths of the strata larger for lower packing fraction (Fig. 4, $\phi \approx 0.25$) and, conversely, the modulation is greater and the strata widths smaller for larger packing fraction (Fig. 6, $\phi \approx 0.70$).

**B. Longitudinal Pair Correlation Functions**

The colloid liquids we have studied, confined to q1D ribbons that support multiple strata, necessarily have anisotropic pair correlation functions. Therefore, we have only examined the pair correlations parallel to the walls within strata.

Figures 7, 8 and 9 display $g_2(x)$ for each of the systems with transverse density distributions displayed in Figs. 4, 5 and 6, for each of the strata in the channel. The $g_2(x)$ for the individual strata have been shifted vertically for clarity. The in-stratum pair correlation functions of the systems with packing fractions $\phi \approx 0.25$ and $\phi \approx 0.41$ are unexceptional, and show very weak dependence of the correlation function on layer distance from the wall despite the differences in stratum densities. There is just a hint in the shapes of the second peaks of $g_2(x)$ in the 11 μm, 14 μm, and 20 μm channels with packing fraction $\phi \approx 0.41$ of very slightly enhanced longitudinal order in the layers immediately adjacent to the two walls.

A q2D colloid suspension confined to a planar gap of order $1.2\sigma$ freezes at a packing density of about 0.7. How this freezing density depends on ribbon width, i.e. reduction of the infinite *xy* space to a narrow channel, is not known, but in our high density ribbon samples the density distributions look ordered, essentially the same along lines that are parallel to and canted relative to the walls, and hexagonally isotropic on the scale length of the channel width, as shown by the images in Fig. 2b, and the particle trajectories (lower panel in Fig. 3). For these systems we calculated both $g_2(x)$ (Fig. 9) and $g_2(r)$ (Fig. 10), and we take the latter to be meaningful for $r < w_{eff}/2$, shown in the shaded regions.

The q1D pair correlation functions $g_2(x)$ of the systems show a marked dependence of the peak widths on channel width; the peaks are narrowest for the 20 μm channel and widest for the 14 μm channel (Fig. 9e), unlike the behavior of the peaks of $g_2(x)$ in systems with smaller packing fractions. This behavior is analogous to that found in the very high-density regime in simulations of hard spheres confined to ribbon channels by smooth hard walls [18]. In that regime, if $w/\sigma$ supports an integer number of strata (i.e. a commensurate configuration) the system packs in a perfect 2D triangular lattice. If such a condition is satisfied, then $\chi$, defined as $\chi \equiv 1 + 2(w-\sigma)/\sqrt{3}\sigma$, is an integer [18].

As shown in Table 1, the nominal $\chi$ values for our high-density systems are all close to an integer, yet the uncertainty introduced from the particle diameter makes in unrealistic to determine whether the width of the peaks in $g_2(x)$ is associated with the closeness of $\chi$ to an integer. However, this lattice has zero shear modulus, hence can flow like a liquid parallel to the walls. If there is a little extra space between the walls, and $w/\sigma$ supports somewhat more than an integer number of strata, there is a buckling instability in the $y$-direction, the system breaks into many triangular solid regions that are displaced in the $x$-direction, and the overall effect is an increase in the average widths of the strata.

At packing fraction $\phi \approx 0.7$, even though the peaks of $g_2(x)$ are narrow and not overlapping, they decay with increasing $x$, clearly indicating that the order is that of a liquid [13]. On the other hand, the pair correlation function $g_2(r)$ shown in the shaded regions in Fig. 10 closely resembles that of a q2D colloid crystal at a similar packing fraction [15]. The differences arise from the inclusion of inter-stratum pair separations in $g_2(r)$ that are obviously not contained in $g_2(x)$.

We note that the first peaks of $g_2(x)$ are not located at $x = \sigma$, even though the silica spheres used are essentially charge neutral. There are likely two dominant contributions to this shift, polydispersity of particle size and a weak attraction between colloid particles. Although the nominal polydispersity specified by the manufacturer is 3%, the apparent polydispersity is much higher (see Fig. 2a), and the effect of the polydispersity on $g_2(x)$ is to shift the first peak to larger $x$ [19]. In addition, we have shown previously that the confined colloid suspensions we have studied, which do not wet the walls, behave as if there is an effective weak colloid-colloid attraction (see Appendix for further discussion of the optical artifact correction for the confinement-induced effective pair-potential).

### C. Phase Shift in the Stratified Longitudinal Pair Distribution Function

A more interesting and striking density dependence of $g_2(x)$ is found when one compares this function in slices taken at the maxima and minima of the transverse density distribution. We show the results of this comparison in Fig. 11. Consider the peak of the density stratum adjacent to a wall and the minimum between that stratum and the next stratum. The first peaks of $g_2(x)$ for the slices through the density maximum of the first stratum and through the minimum between the first two strata have the same location, but the data reveal a phase shift in the locations of the second peaks of $g_2(x)$. A smaller phase shift is found between these peaks of $g_2(x)$ for slices through the maximum density of the second stratum and through the minimum in density between the second and third strata. This phase shift resembles that found by Kjellander and Sarman [20] for the same slices of the in-plane pair correlation function of a hard sphere fluid in a planar slit with hard walls with separations $2.1\sigma$ and $15.2\sigma$. Indeed, the magnitude of the shift, $0.2\sigma$, is the same as that found from the simulation.

### IV. Discussion

Overall, our experimental results illustrate how the properties of a fluid confined to a q1D ribbon channel are determined by the competition between fluctuations that prevent ordering and the ordering that would be generated by the most favorable exploitation of the free volume of the system. Specifically, our experimental studies confirm the predictions of the three principal properties of colloid liquids confined to occupy a q1D ribbon that follow from that competition. These are (1) that the density distribution of the liquid transverse to the ribbon channel exhibits stratification; (2) that even at the highest density the order along the strata, as measured by the longitudinal pair correlation function, is characteristic of a liquid; and (3) that the q1D pair correlation function in different strata exhibits anisotropic behavior, resembling that found in a Monte Carlo simulation of the in-plane pair correlation function of a hard sphere fluid in a planar slit [20].

The phase shift between the longitudinal pair correlations in slices containing the density maximum of the first stratum and the density minimum between the first two strata of a hard sphere fluid contained between smooth hard walls is correctly predicted using an anisotropic version of the Percus-Yevick integral equation. Kjellander and Sarman [20] interpret this structural feature as a consequence of the stronger competition for free volume between a particle in the inter-stratum space and those in the adjacent dense strata than between particles in the inter-stratum space. This interpretation is not consistent with the lack of phase shift between the longitudinal pair correlations in slices containing the density maximum of, say, the third stratum and the density minimum between the third and fourth strata (see Fig. 11b).

The observed independence of density of the longitudinal pair correlation function across the internal strata of the density distribution is like that found for the stratified liquid-vapor interface of a metal, for example Cs. Harris and Rice [21] showed that this behavior may be understood using the Fisher-Methfessel (FM) formulation of the local density approximation. Fisher and Methfessel [22] introduced the assumption that the pair correlation function of a 3D inhomogeneous fluid of hard spheres can be approximated by the pair correlation function of a homogeneous hard sphere fluid at a nominal density defined by averaging the point density over a sphere with diameter equal to the hard sphere diameter and centered at the point of contact of the particles. This conceptual picture can be extended to the q2D and q1D cases with obvious definitions of the domain over which the averaging is to be carried out. The FM approximation has the effect of keeping the lowest order representation of the force necessary to maintain the density inhomogeneity in the liquid, unlike the conventional local density approximation that completely neglects that force.

Examination of the density distribution shown in Fig. 11a shows that the average density defined by the FM procedure appropriate to the geometry varies relatively little across the space occupied by the interior strata, hence so will the longitudinal pair correlation function. We suggest that the observed phase shift between the longitudinal pair correlations in slices containing the density maximum of the first stratum and the density minimum between the first two strata is a consequence of different amplitudes of motion perpendicular to the walls of the particles in the stratum adjacent to the wall and the next stratum (see Fig. 3 middle panel), which generates an anisotropic environment for any particles in the density minimum region.

## V. Acknowledgments

The artifact correction described in the Appendix was carried out with the help of the IDL codes developed by Dr. Poli and Prof. Grier. We are grateful for their assistance in the application to our imaging data. The research reported in this paper was supported by a Dreyfus Foundation Mentor Grant to SAR and by the NSF funded MRSEC Laboratory at The University of Chicago (NSF DMR-0213745).

Table 1. Channel widths; $w$ refers to the width of the grooves on the master mold and $w_{eff}$ to the effective width calculated as described in the text. For higher packing fractions, $\phi \approx 0.7$, $\chi$ is also calculated (note that the value of $\sigma = 3.2\,\mu m$, determined from the position of the first peak in pair-distribution functions, is used in calculating $\chi$).

| $\sigma$ (μm) | $w$ (μm) | $w_{eff}$ (μm) | $w_{eff}/\sigma$ | $\chi$ |
|---|---|---|---|---|
| 1.58 | 5.0 | 3.1 | 2.0 | |
| 1.58 | 8.0 | 5.7 | 3.6 | |
| 1.58 | 11.0 | 8.9 | 5.6 | |
| 1.58 | 14.0 | 11.6 | 7.4 | |
| 1.58 | 20.0 | 17.7 | 11.3 | |
| 3.01±0.2 | 8.0±0.1 | 6.2±0.2 | 2.1±0.2 | 2.1±0.1 |
| 3.01±0.2 | 14.0±0.1 | 11.4±0.2 | 3.8±0.3 | 3.9±0.3 |
| 3.01±0.2 | 20.0±0.1 | 17.6±0.2 | 5.8±0.4 | 6.1±0.4 |
| 3.01±0.2 | 25.0±0.1 | 22.5±0.2 | 7.4±0.5 | 7.9±0.6 |

# VI. Appendix

As reported in earlier work, we have found that in q1D and q2D colloid suspensions that do not wet the confining walls there is an effective colloid-colloid interaction with weak attractive well. In this Appendix we address the suggestion, originally raised by Bechinger and coworkers [23, 24] and later also considered by several research groups, that an optical artifact arising from the overlap of the images of colloid particles with small separation generates an error in the measurements of the separation of the particles that results in fictitious features in the pair-interaction potential. Specifically, they identified a systematic deviation between the digitized (or apparent) particle separation, $\tilde{r}$, and the real one, $r$ ($\tilde{r} < r$) when particles are close to contact, caused by optical image overlap. To ascertain the influence of this artifact on the inference that in the q1D systems we have studied there is an effective colloid-colloid interaction with weak attractive well, we have revisited our data analysis for the q1D colloid liquid and explicitly accounted for the artifact correction.

Several schemes for the correction of the referred to image distortion have been proposed and applied to remove the error in the calculation of particle pair-interaction potential. We chose to use one of the methods introduced by Polin et al [25] that allows us to correct the artifact retrospectively with the imaging data used previously to calculate the pair potential. Briefly, one first identifies a sphere in an image in which all particles are very well separated from each other, so there can be no image overlaps. Then the image of such a sphere is cropped and a two-sphere image is constructed by duplicating the cropped image a designated distance $r$ away from the original (cropped) image; this process is repeated for variable $r$. The apparent separation, $\tilde{r}$, between the two spheres in the constructed image is determined using the standard method [17].

Finally, the difference between the apparent separation and the real separation, $\Delta r = \tilde{r} - r$, is used to obtain the undistorted pair distribution function, $g_2(r)$, from the distorted pair distribution function, $g_2(\tilde{r})$, with the relation $g_2(r)dr = g_2(\tilde{r})d\tilde{r}$, which can be approximated to $g_2(r) = g_2(r + \Delta r)\left[1 + \frac{d}{dr}\Delta r\right]$ at low density (Eq. (4) in [25]).

Figure A.1 shows a snap shot of our q1D colloid liquid at a q1D packing fraction $\eta=0.18$. The sphere in the middle of the group was selected to construct the two-sphere composite, and 100 frames of such images were used to assess the importance of the optical artifact correction. Figure A.2 shows the relation between the difference $\Delta x = \tilde{x} - x$ and the real separation $x$. The pair correlation function and effective pair-interaction potential, before and after the correction, are plotted in Figs. A.3 and A.4. Clearly the correction process has resulted in the removal of some seemingly non-physical features in the $g_2(\tilde{x})$ at this density, such as the excessive sharpness of the first peak and the lack of the first minimum. The effect of the corrections is to slightly shift the repulsive part of the interaction potential to smaller $x$, but the depth of the attractive well is not significantly changed within the experimental precision (the oscillations in the data curve).

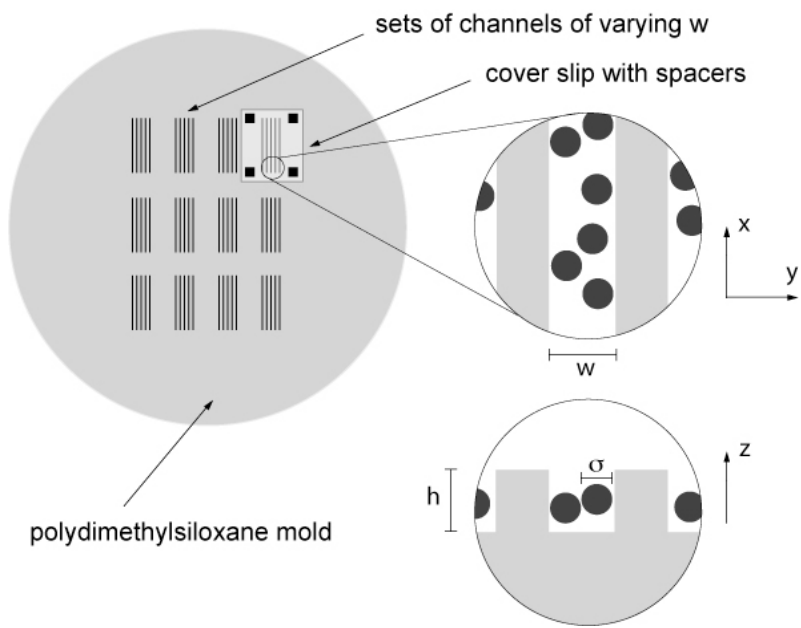

Figure 1. Schematic diagram of the experimental system.

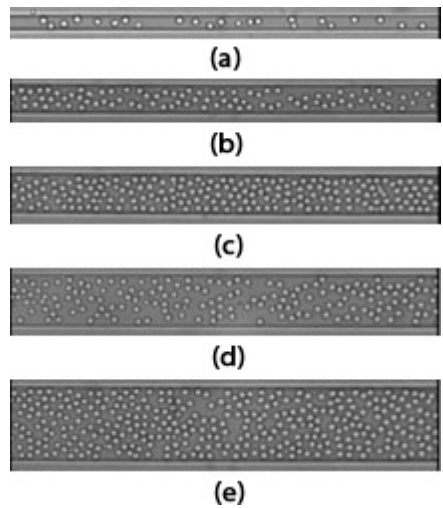

Figure 2a. Images of different q1D ribbon systems, all with colloid diameter 1.58 μm: (a) w = 5 μm, $\phi$ = 0.185; (b) w = 8 μm, $\phi$ = 0.260; (c) w = 11 μm, $\phi$ = 0.395; (d) w = 14 μm, $\phi$ = 0.440; (e) w = 20 μm, $\phi$ = 0.322. The widths listed correspond to the master mold without any adjustment for effect of non-wetting of the walls by the aqueous colloid suspension.

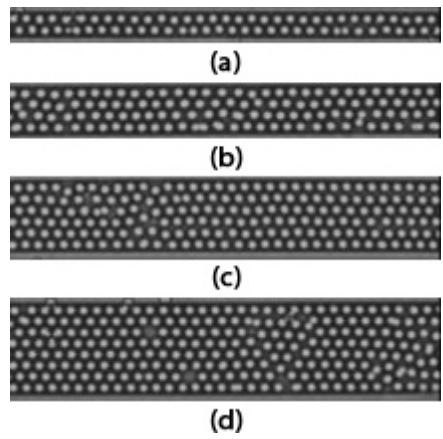

Figure 2b. Images of different q1D ribbon systems, all with colloid diameter 3.01 μm: (a) w = 8 μm, $\phi$ = 0.686; (b) w = 14 μm, $\phi$ = 0.692; (c) w = 20 μm, $\phi$ = 0.702; (d) w = 25 μm, $\phi$ = 0.697. The widths listed correspond to the master mold without any adjustment for effect of non-wetting of the walls by the aqueous colloid suspension.

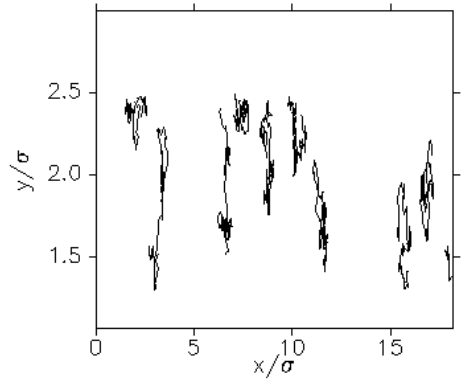

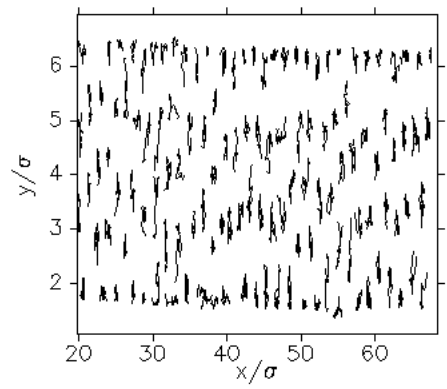

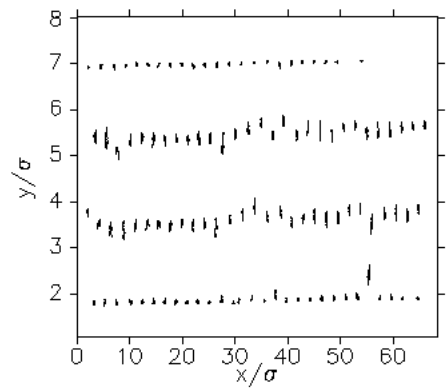

Figure 3. Sample trajectory images. From top to bottom: $\sigma = 1.58$ μm, $w = 5$ μm, $\phi = 0.185$; $\sigma = 1.58$ μm, $w = 11$ μm, $\phi = 0.395$; $\sigma = 3.01$ μm, $w = 14$ μm, $\phi = 0.692$. Both axes are normalized to measure distance in sphere diameters.

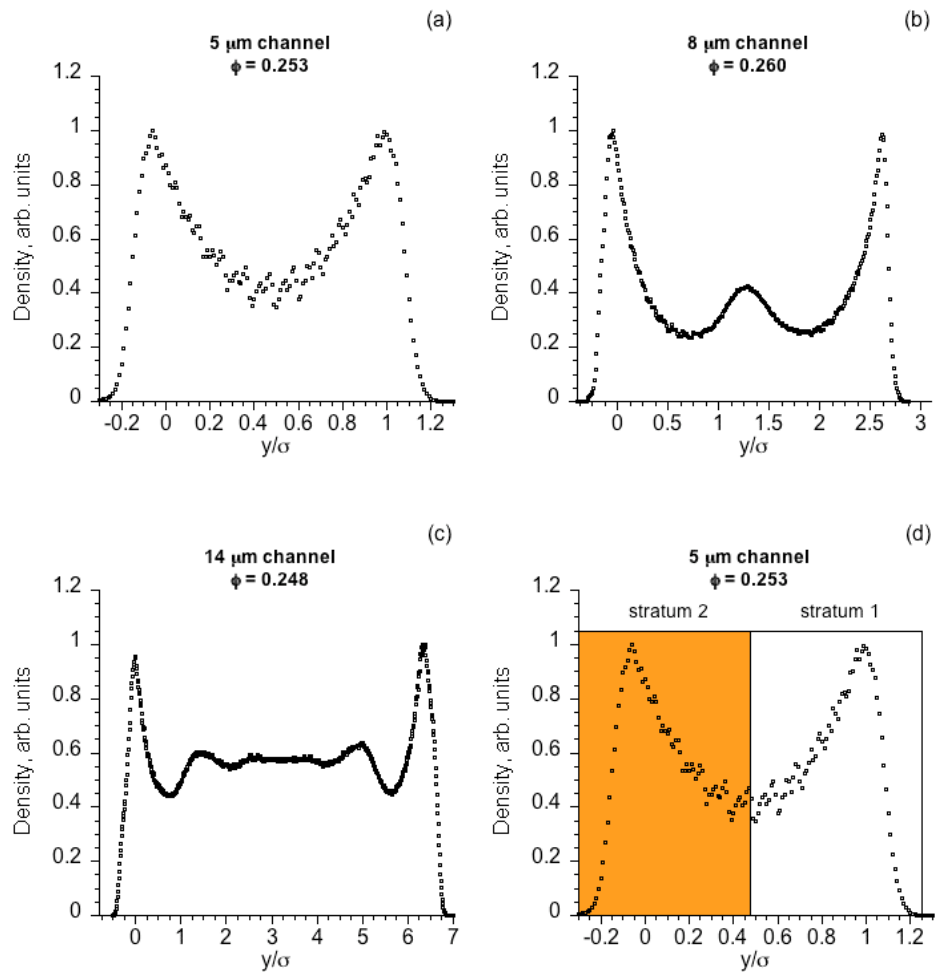

Figure 4. Transverse density distributions of 1.58 μm particles with packing fraction $\phi \approx$ 0.25 in channels with various widths: (a) $w = 5$ μm, (b) $w = 8$ μm, (c) $w = 14$ μm. Part (d) shows a sample of how the data were selected to compute $g_2(x)$ for Fig. 7-9.

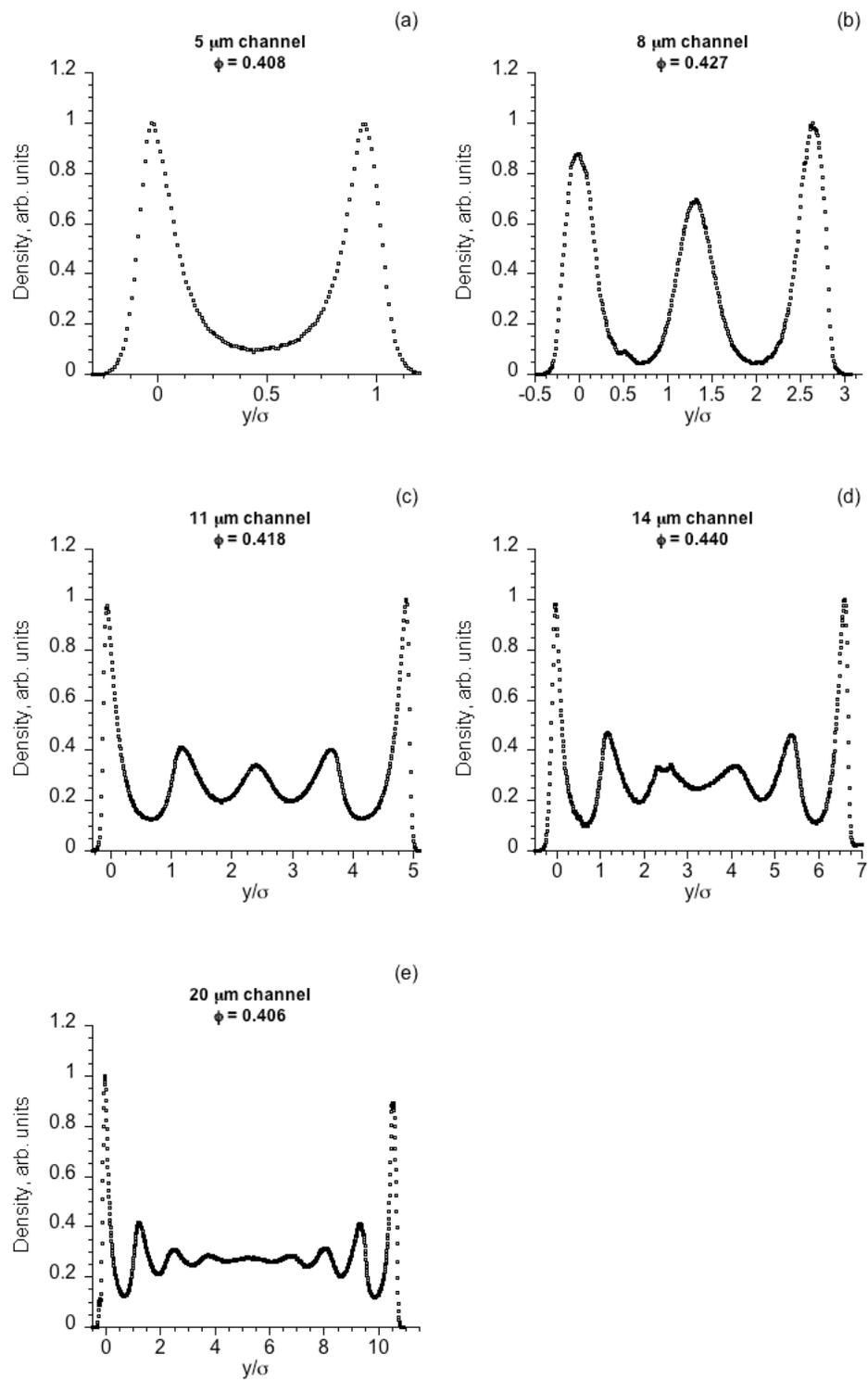

Figure 5. Transverse density distributions of 1.58 μm particles with packing fraction $\phi \approx$ 0.41 in channels with various widths: (a) $w = 5$ μm, (b) $w = 8$ μm, (c) $w = 11$ μm, (d) $w = 14$ μm, (e) $w = 20$ μm.

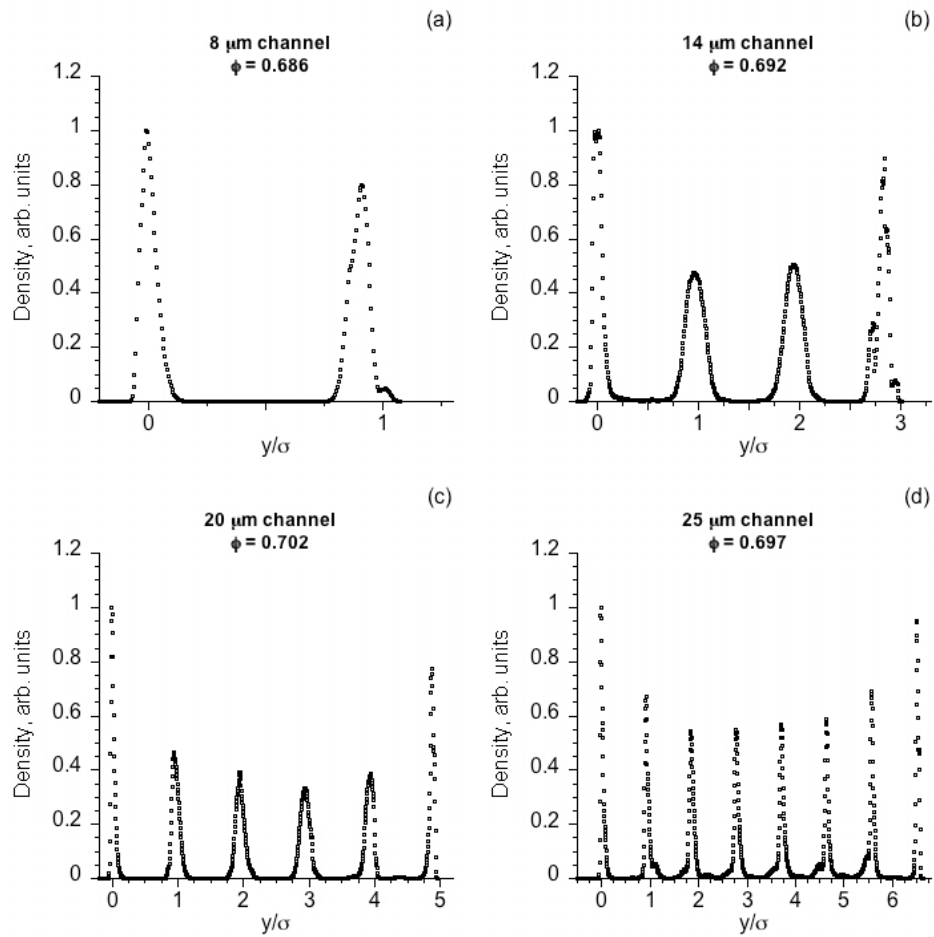

Figure 6. Transverse density distributions of 3.01 μm particles with packing fraction $\phi \approx$ 0.70 in channels with various widths: (a) $w = 8$ μm, (b) $w = 14$ μm, (c) $w = 20$ μm, (d) $w = 25$ μm.

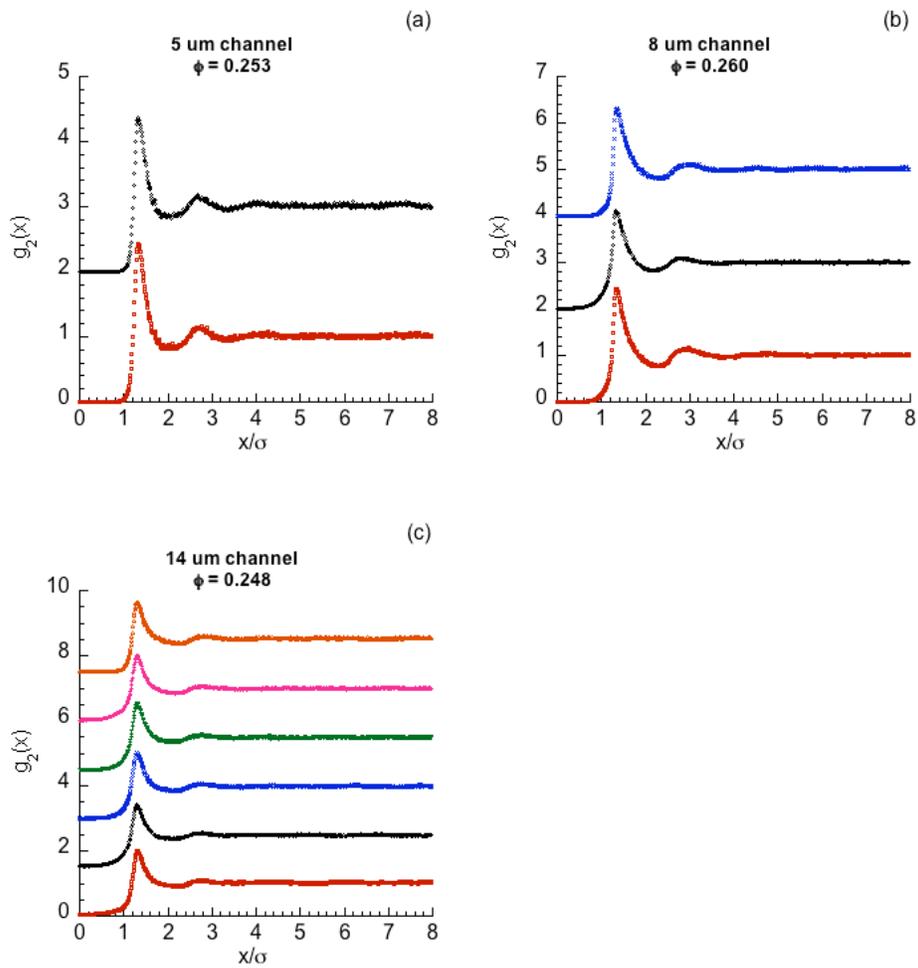

Figure 7. Pair correlation functions along individual strata of 1.58 μm particles with packing fraction $\phi \approx 0.25$ in channels with various widths: (a) $w$ = 5 μm, (b) $w$ = 8 μm, (c) $w$ = 14 μm, corresponding to density profiles in Fig 4. For each stratum, $g_2(x)$ was shifted vertically for clarity, with the top-most graph corresponding to the left-most stratum in density distribution.

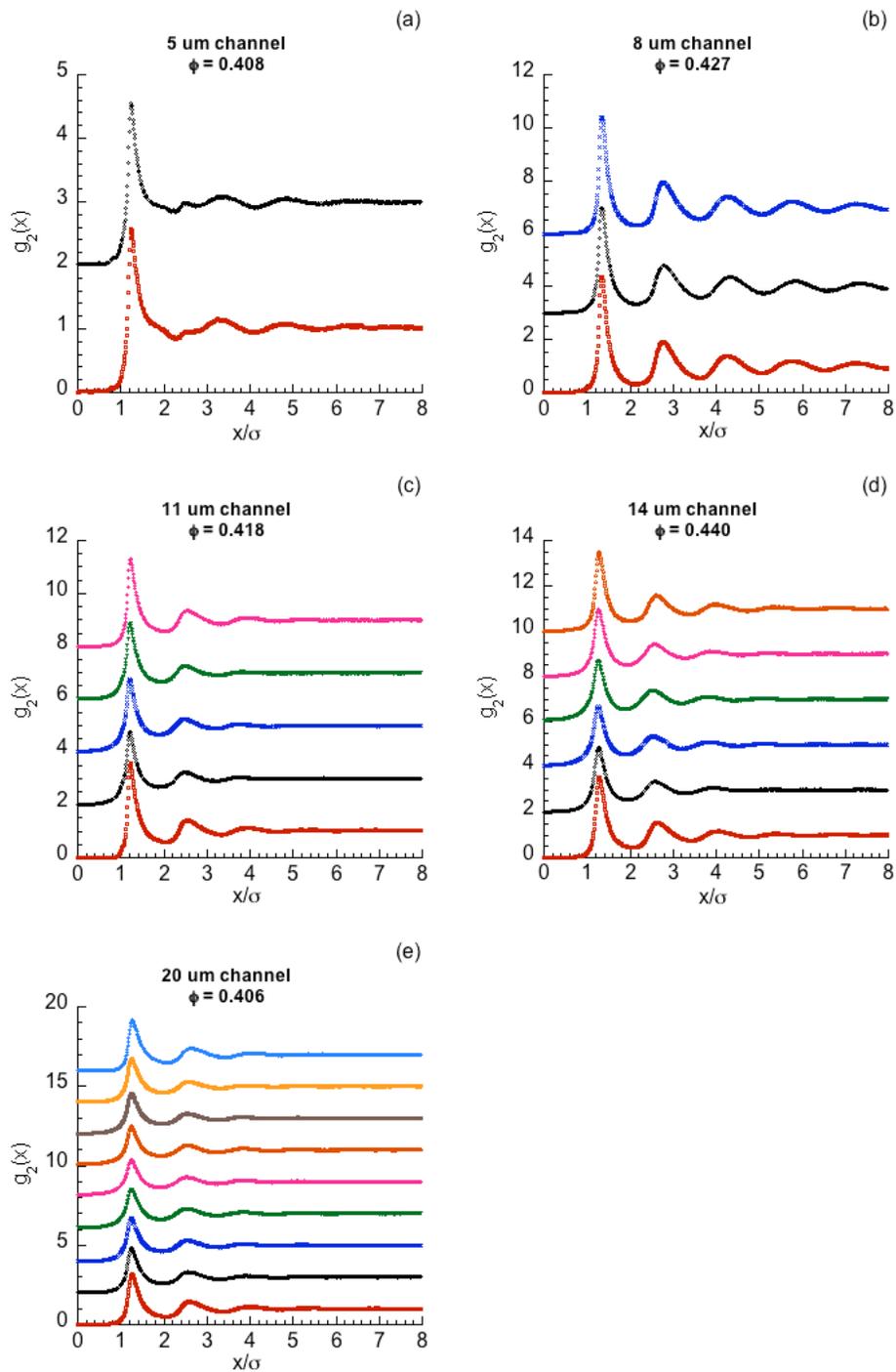

Figure 8. Pair correlation functions along individual strata of 1.58 μm particles with packing fraction $\phi \approx 0.41$ in channels with various widths: (a) $w = 5$ μm, (b) $w = 8$ μm, (c) $w = 11$ μm, (d) $w = 14$ μm, (e) $w = 20$ μm, corresponding to density profiles in Fig 5. For each stratum, $g_2(x)$ was shifted vertically for clarity, with the top-most graph corresponding to the left-most stratum in density distribution.

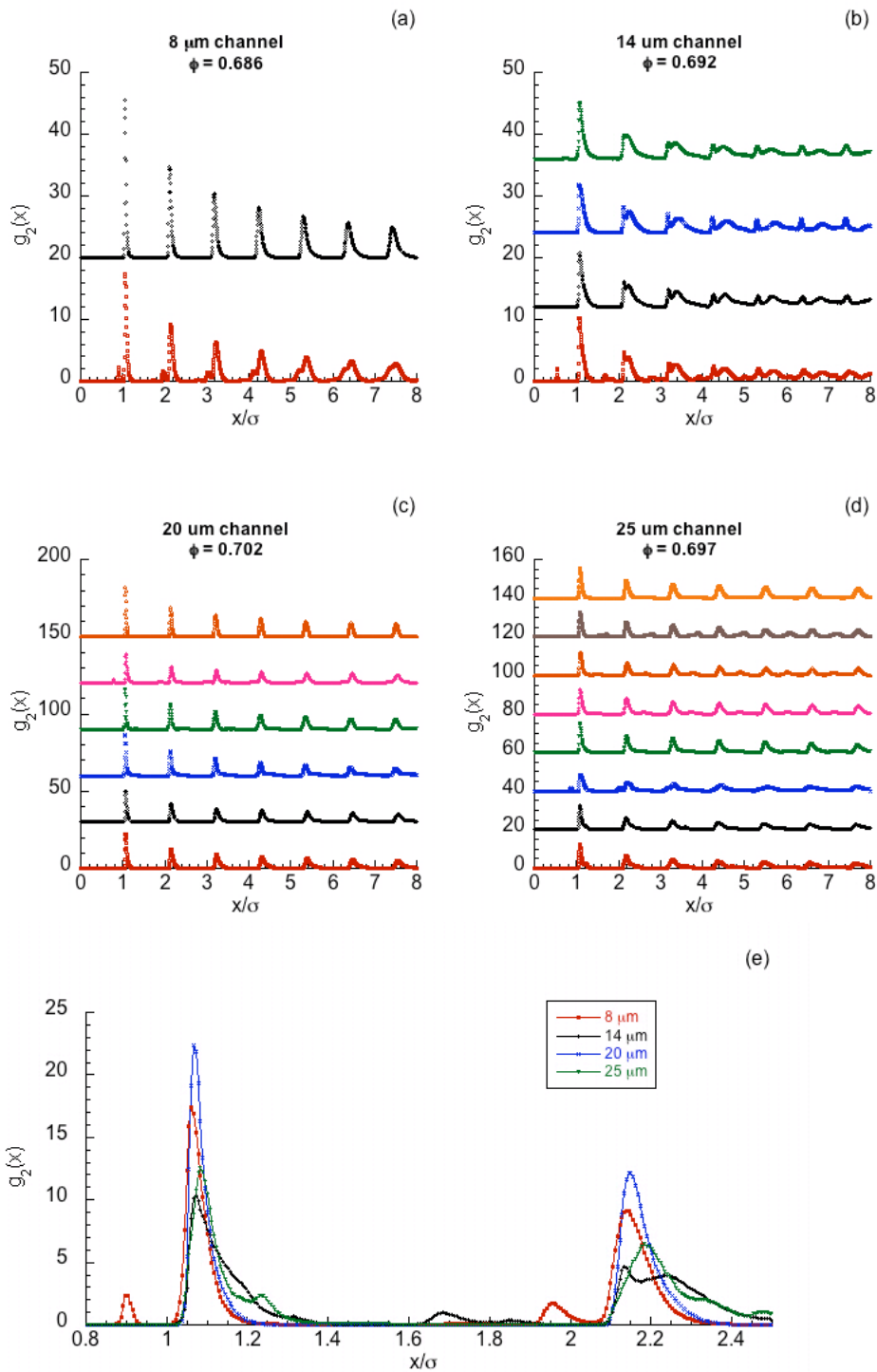

Figure 9. Pair correlation functions along individual strata of 3.01 µm particles with packing fraction $\phi \approx 0.70$ in channels with various widths: (a) $w = 8$ µm, (b) $w = 14$ µm, (c) $w = 20$ µm, (d) $w = 25$ µm, corresponding to density profiles in Fig 6. For each stratum, $g_2(x)$ was shifted vertically for clarity, with the top-most graph corresponding to the left-most stratum in density distribution. Part (e) compares peak widths for the first stratum.

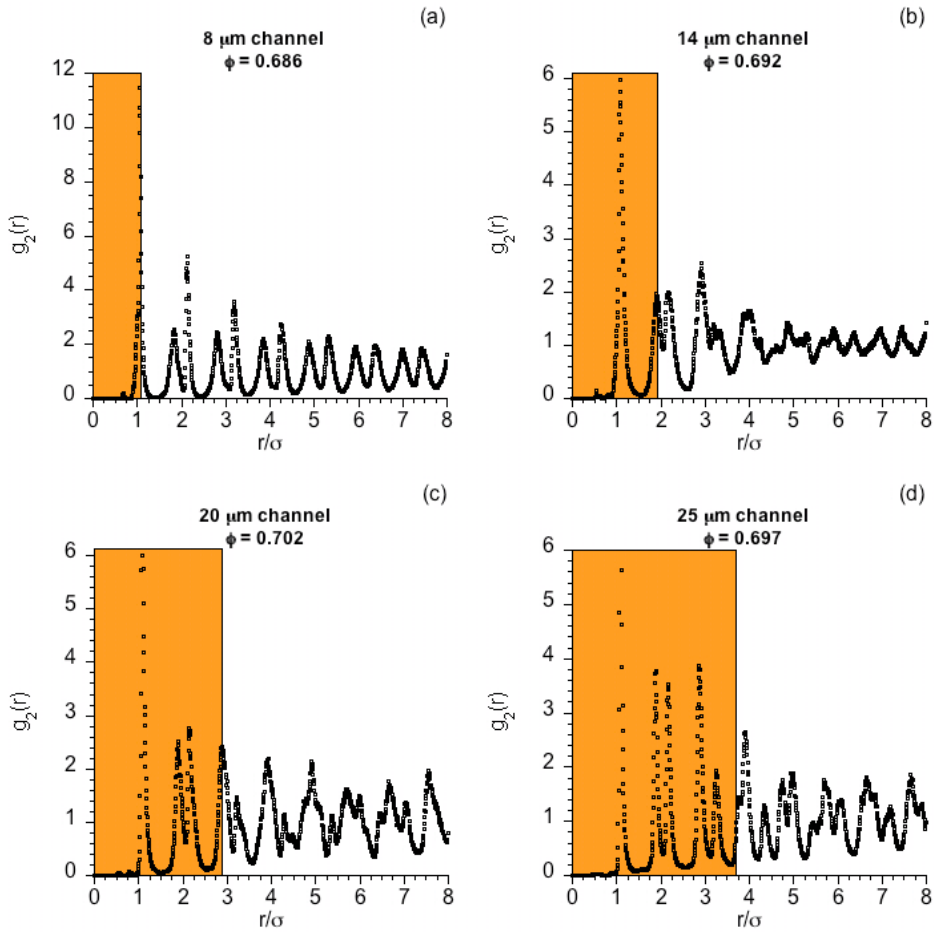

Figure 10. Pair correlation functions, $g_2(r)$, of 3.01 μm particles with packing fraction $\phi \approx 0.70$ in channels with various widths: (a) $w = 8$ μm, (b) $w = 14$ μm, (c) $w = 20$ μm, (d) $w = 25$ μm. The shaded regions indicate $r < w_{eff}/2$, where $g_2(r)$ is taken to be meaningful.

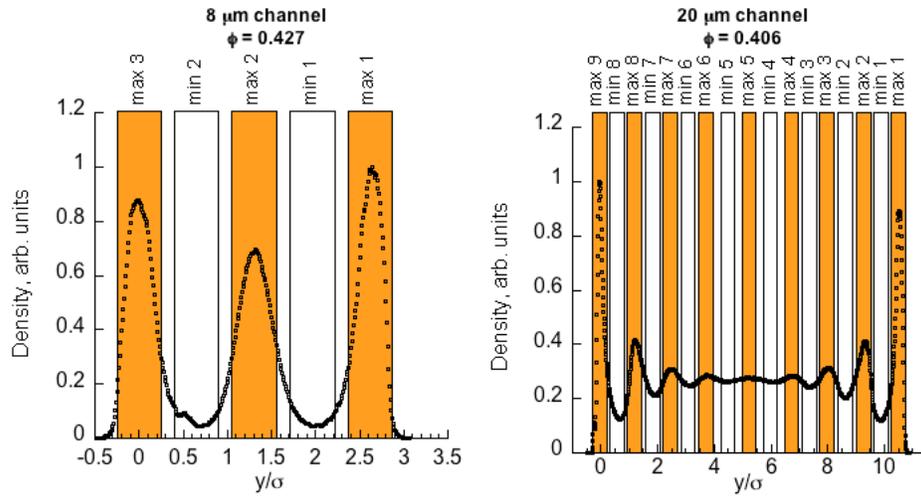

Figure 11a. Slices of the transverse density distribution function in which $g_2(x)$ of Fig. 11b was computed. Width of each slice is $0.5\sigma$.

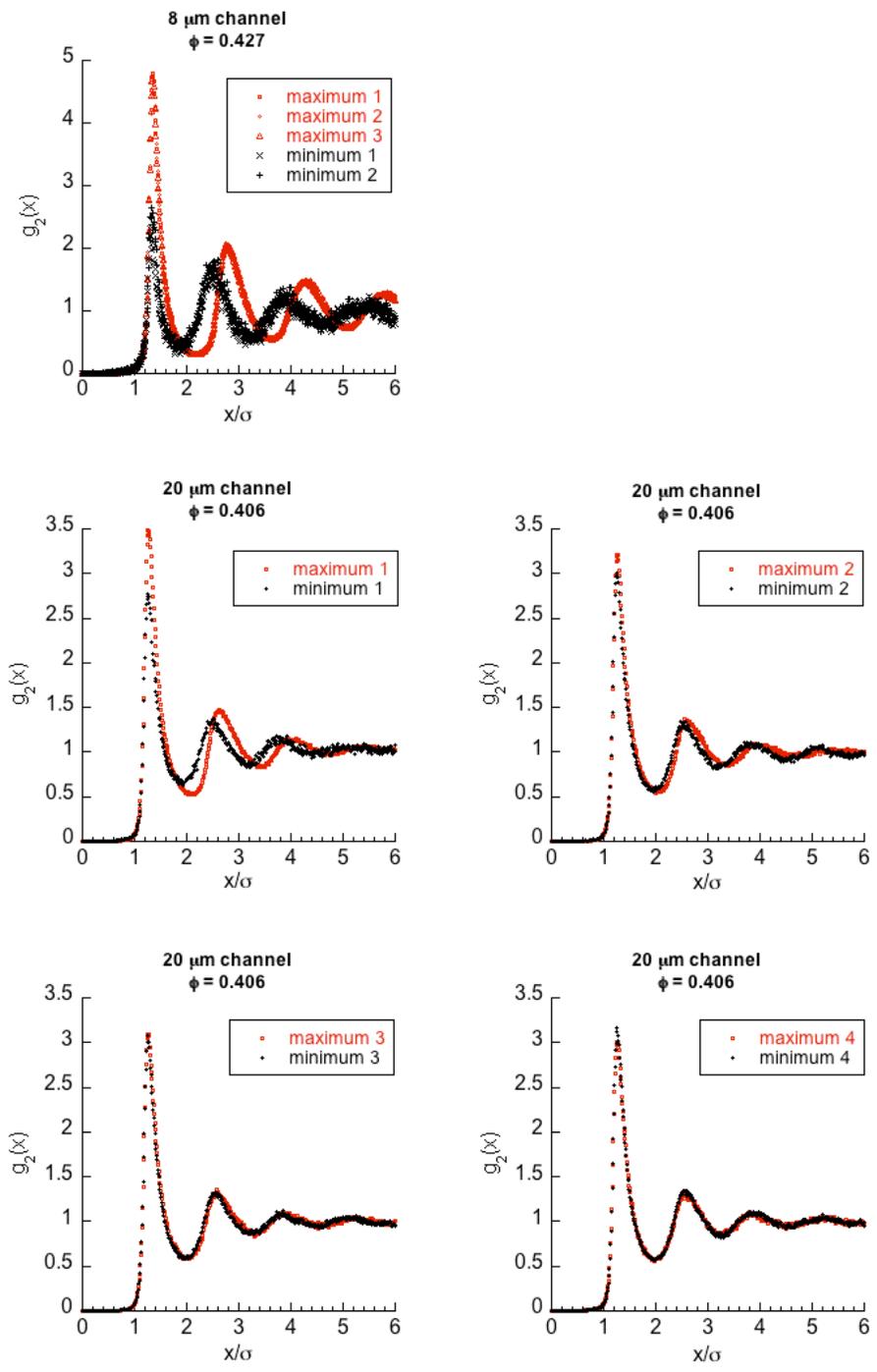

Figure 11b. Longitudinal pair correlation functions $g_2(x)$ at various positions along the transverse density distribution.

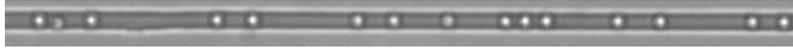

Figure A.1. A snap shot of q1D colloid liquid at a q1D packing fraction $\eta=0.18$ used for image correction.

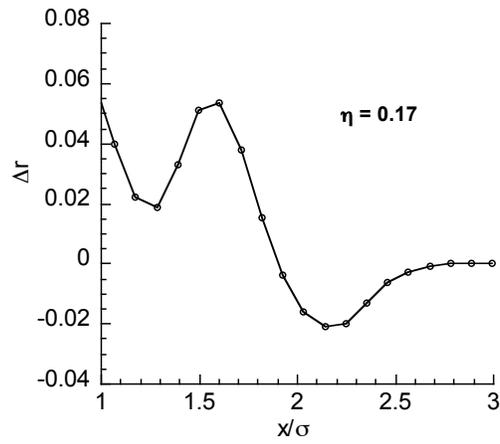

Figure A.2. Relation between the difference $\Delta x = \tilde{x} - x$ and the real separation $x$.

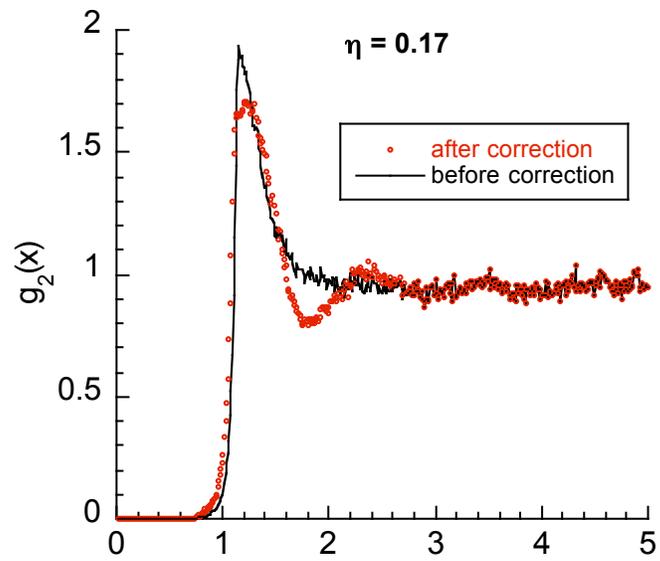

Figure A.3. Pair correlation function before and after correction.

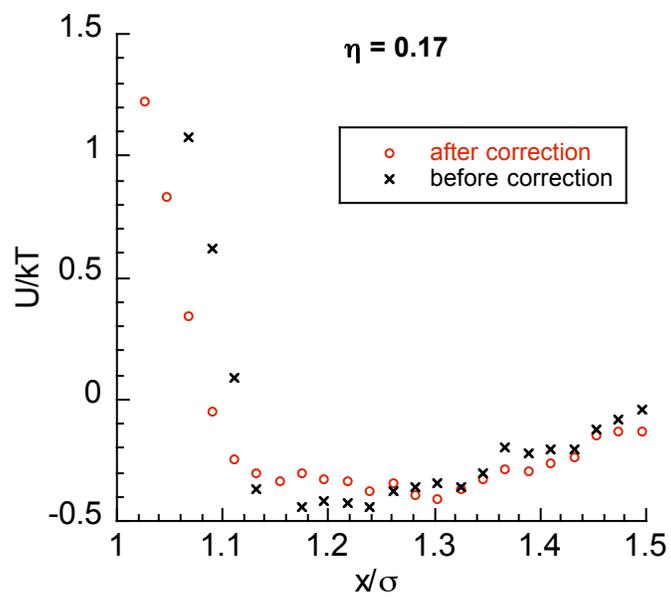

Figure A.4. Effective pair-interaction potential before and after correction.